\documentclass[
 aip,
 amsmath,
 amssymb,
 reprint,
]{revtex4-1}

\usepackage{graphicx}
\usepackage{dcolumn}
\usepackage{bm}
\usepackage{xcolor}
\usepackage{soul}

\begin{document}
	
	\preprint{AIP/123-QED}
	
	\title[]{Intra- and Inter-Conduction Band Optical Absorption Processes in $\beta$-Ga$_2$O$_3$}
	
	\author{Arjan Singh*}
	\email{as2995@cornell.edu}
	
	\author{Okan Koksal*}
	\affiliation{School of Electrical and Computer Engineering, Cornell University, Ithaca, NY 14853, USA.}
	
	\author{Nicholas Tanen}
	\affiliation{Department of Materials Science and Engineering, Cornell University, Ithaca, NY 14853, USA.}
	
	\author{Jonathan McCandless}
	\affiliation{School of Electrical and Computer Engineering, Cornell University, Ithaca, NY 14853, USA.}
	
	\author{Debdeep Jena}
	
	\author{Huili (Grace) Xing}
	\affiliation{School of Electrical and Computer Engineering, Cornell University, Ithaca, NY 14853, USA.}
	\affiliation{Department of Materials Science and Engineering, Cornell University, Ithaca, NY 14853, USA.}
	
	\author{Hartwin Peelaers}
	\affiliation{Department of Physics and Astronomy, University of Kansas, Lawrence, KS 66045, USA.}
	
	\author{Farhan Rana}
	\affiliation{School of Electrical and Computer Engineering, Cornell University, Ithaca, NY 14853, USA.}
	
	\date{\today}

	\begin{abstract}
		$\beta$-Ga$_2$O$_3$ is an ultra-wide bandgap semiconductor and is thus expected to be optically transparent to light of sub-bandgap wavelengths well into the ultraviolet.  Contrary to this expectation, it is found here that free electrons in n-doped $\beta$-Ga$_2$O$_3$  absorb light from the IR to the UV wavelength range via intra- and inter-conduction band optical transitions. Intra-conduction band absorption occurs via an indirect optical phonon mediated process with a $1/\omega^{3}$ dependence in the visible to near-IR wavelength range. This frequency dependence markedly differs from the $1/\omega^{2}$ dependence predicted by the Drude model of free-carrier absorption. The inter-conduction band absorption between the lowest conduction band and a higher conduction band occurs via a direct optical process at $\lambda \sim 349$ nm (3.55 eV). Steady state and ultrafast optical spectroscopy measurements unambiguously identify both these absorption processes and enable quantitative measurements of the inter-conduction band energy, and the frequency dependence of absorption. Whereas the intra-conduction band absorption does not depend on light polarization, inter-conduction band absorption is found to be strongly polarization dependent. The experimental observations, in excellent agreement with recent theoretical predictions for $\beta$-Ga$_2$O$_3$, provide important limits of sub-bandgap transparency for optoelectronics in the deep-UV to visible wavelength range, and are also of importance for high electric field transport effects in this emerging semiconductor.
	\end{abstract}
	
	
	
	\maketitle
	
	The ultra wide-bandgap semiconductor $\beta-$Ga$_2$O$_3$ is expected to be transparent to light in the IR to UV wavelength range because of its large 4.4-4.9 eV energy bandgap \cite{ricci_bg, masahiro_bg, matsumoto_bg, tippins_bg}. This transparency, combined with the ability to grow large area single-crystal $\beta$-Ga$_2$O$_3$ substrates \cite{aida_efg, kuramata_efg, galazga_czochralski}, makes $\beta$-Ga$_2$O$_3$ a promising material for solar-blind deep-UV photodetectors and photovoltaic cells \cite{hu_uvdet, ji_uvdet, oshima_uvdet, minami_solarcell}. Free-electrons in the conduction band are needed to maintain the required conductivity in such optoelectronic devices. However, free-electrons also enable optical absorption at photon energies below the bandgap, limiting the material's transparency. 
	
	Sub-bandgap light absorption by free carriers in $\beta$-Ga$_2$O$_3$ can occur by two distinct transition processes. First, electrons in the lowest conduction band can absorb a photon, and undergo a direct transition to a higher conduction band (inter-conduction band absorption). Such optical transitions have been reported previously in narrow-bandgap group III-V semiconductors \cite{eijiro_intercb_III_V, glosser_intercb_insb}, heavily doped n-type Silicon \cite{balkanski_intercb_silicon}, and also in heavily Fluorine doped Tin Oxide (FTO) \cite{canestraro_intercb_FTO}. Second, electrons in the lowest conduction band can also absorb a photon and transition to higher energy states within the same conduction band (intra-conduction band absorption). These latter transitions require a momentum conserving process, such as phonon emission/absorption by the electron or electron-ionized impurity scattering \cite{baker-finch_sifc, perkowitz_gaasfc1, perkowitz_gaasfc2, von-baltz_qmfc, peelaers_sno2fc}. In addition to its relevance to optoelectronic devices that rely on the optical transparency of this material, understanding the mechanisms of direct and indirect optical absorption by electrons in the conduction band of $\beta$-Ga$_2$O$_3$ has direct implications on high electron energy transport phenomena, such as impact ionization and high-electric field breakdown voltages, for which this ultra wide-bandgap semiconductor is considered most attractive.
	
    \begin{figure}[ht]
		\includegraphics[width=0.54\columnwidth]{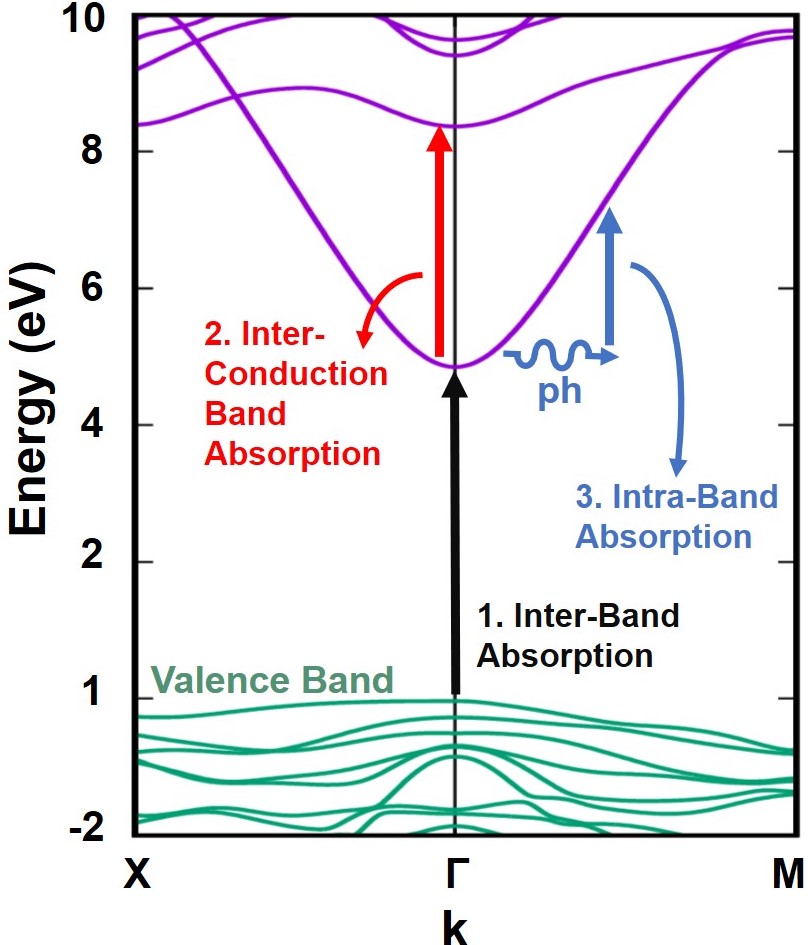}
		\caption{Schematic of inter- and intra-conduction band absorption of light by conduction band electrons in $\beta$-Ga$_2$O$_3$.}
		\label{fig:bands}
	\end{figure}
        
	Peelaers et al. recently calculated the optical absorption cross-sections of free electrons from the first-principle electronic bandstructure of $\beta-$Ga$_2$O$_3$, and predicted unique features of both intra- and inter-conduction band transitions \cite{peelaers_ga2o3fc,peelaers_subbgabs}. In this paper, we report results from steady state and ultrafast optical transmission spectroscopy of doped bulk $\beta$-Ga$_2$O$_3$ substrates, with direct evidence of the predicted intra- and inter-conduction band absorption processes. At the $\Gamma-$point, the lowest conduction band of $\beta$-Ga$_2$O$_3$ has Ga:4s symmetry and the next lowest conduction band has Ga:4p symmetry \cite{dong_optical_defects}. Inter-conduction band optical absorption between these two bands is therefore possible (see Figure \ref{fig:bands}). We observe a strong absorption peak near $\sim 349$ nm (3.55 eV) in doped $\beta$-Ga$_2$O$_3$ samples, which is absent in undoped or lightly doped samples. This is consistent with the predicted inter-conduction band absorption \cite{peelaers_subbgabs}. This absorption feature is found to depend on the polarization of light, and becomes negligibly small for light polarized along the crystal b-axis ([010]).
	
	Additionally, we observe that the strength of the optical absorption in the entire visible to IR wavelength range varies with the photon frequency $\omega$, as $1/\omega^{3}$. This differs from the typical $1/\omega^{2}$ dependence explained by the Drude model for intra-band free-carrier absorption \cite{baker-finch_sifc, von-baltz_qmfc}. Peelaers et al. predicted that indirect longitudinal optical (LO) phonon mediated intra-band optical transitions in $\beta$-Ga$_2$O$_3$ should follow this unconventional $1/\omega^{3}$ dependence in the free carrier absorption coefficient in the visible to the near-IR wavelength range\cite{peelaers_ga2o3fc, peelaers_sno2fc, Peelaers2019b}. Our experimental observations confirm this second prediction as well.

\begin{figure}[ht]
		\includegraphics[width=0.77\columnwidth]{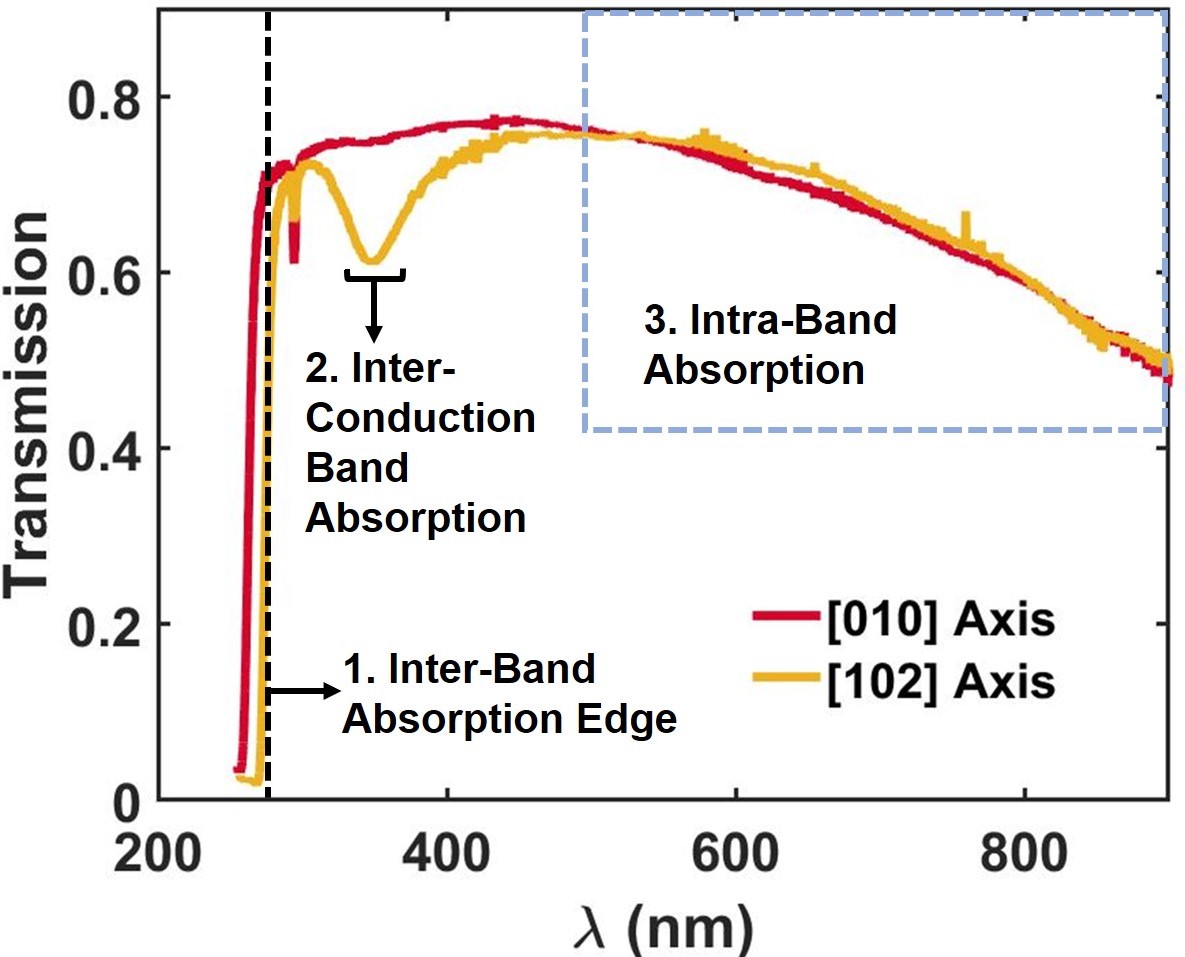}
		\caption{Measured optical transmission through a 300 $\mu$m thick, Sn-doped ($n\sim7\times10^{18}$ cm$^{-3}$), ($\bar{2}01$) $\beta$-Ga$_2$O$_3$ sample for light polarized along the [010] and [102] axes ($T$ = 300K). The data shows prominent features related to inter- and intra-conduction band absorption.}
		\label{fig:trans_pol_dep}
	\end{figure}

	Optical transmission spectroscopy measurements were performed on Sn-doped bulk ($\bar{2}01$) $\beta$-Ga$_2$O$_3$ samples, grown using the edge-defined film-fed growth (EFG) technique by the Tamura Corporation. The sample had a free-electron density of $n\sim7\times10^{18}$ cm$^{-3}$, and a thickness of $\sim 300$ ${\mu}$m. The optical transmission spectrum was measured at room temperature using light polarized along different crystal axes of $\beta$-Ga$_2$O$_3$. Figure \ref{fig:trans_pol_dep} shows the measured optical transmissivity as a function of the light wavelength for light polarized along the [102] and [010] crystal axes. The valence band to lowest conduction band absorption edge is polarization dependent, and is measured to be at $\sim$260 nm (4.77 eV) and $\sim$273 nm (4.54 eV) for light polarized along the [010] and [102] axes respectively. This is consistent with what has been observed experimentally and calculated theoretically \cite{ricci_bg, peelaers_subbgabs}. In addition to the standard optical absorption edge, the transmission spectra in Figure \ref{fig:trans_pol_dep} shows two distinct free-carrier absorption features below the bandgap.
		
\begin{figure}[ht]
		\includegraphics[width=0.74\columnwidth]{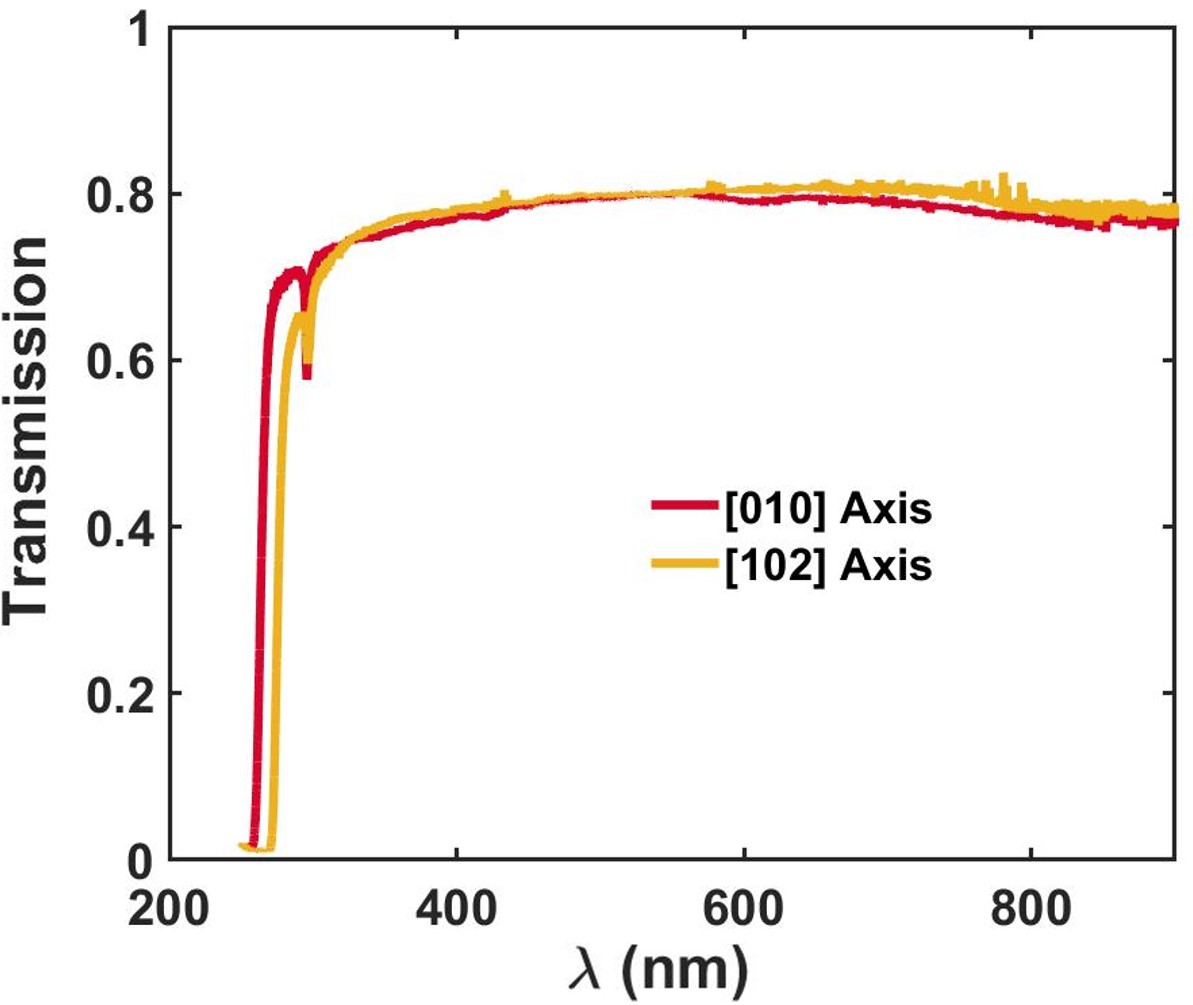}
		\caption{Measured optical transmission through a 300 $\mu$m thick, unintentionally doped ($n \sim 10^{17}$ cm$^{-3}$), ($\bar{2}01$) $\beta$-Ga$_2$O$_3$ sample is plotted for different light polarizations at $T$=300 K. The data shows no features related to inter- or intra-conduction band absorption that are visible in Figure \ref{fig:trans_pol_dep}.}
		\label{fig:undoped_trans}
	\end{figure}

	First, broad wavelength-dependent and polarization-independent absorption is observed in the entire visible to near-IR wavelength range. Second, a clear absorption dip is seen around $\sim$ 349 nm. The strength of this absorption was found to be polarization dependent. Measurements performed on undoped samples did not exhibit either of these two absorption features. For example, Figure \ref{fig:undoped_trans} shows the transmission through a 300 $\mu$m thick, unintentionally doped ($n \sim 10^{17}$ cm$^{-3}$), ($\bar{2}01$) $\beta$-Ga$_2$O$_3$ sample (also grown by the Tamura Corporation using the EFG method). This indicates that these two absorption features shown in Figure \ref{fig:trans_pol_dep} are due to the presence of free carriers. We attribute the broad wavelength-dependent absorption to intra-conduction band absorption and the absorption dip near 349 nm to inter-conduction band absorption. Note that in addition to the free-carrier absorption described, a sharp absorption feature centered at 296 nm is also seen in the transmission spectrum of both doped and undoped ($\bar{2}01$) $\beta$-Ga$_2$O$_3$ samples (see Figures \ref{fig:trans_pol_dep} and \ref{fig:undoped_trans}). While the origin of this absorption is unclear, its presence in both doped and undoped samples suggests that it is not related to doping or to the presence of electrons in the conduction band. In what follows, we first discuss the intra-band absorption spectra and then the inter-conduction band absorption process.
	
\begin{figure}[ht]
		\includegraphics[width=1\columnwidth]{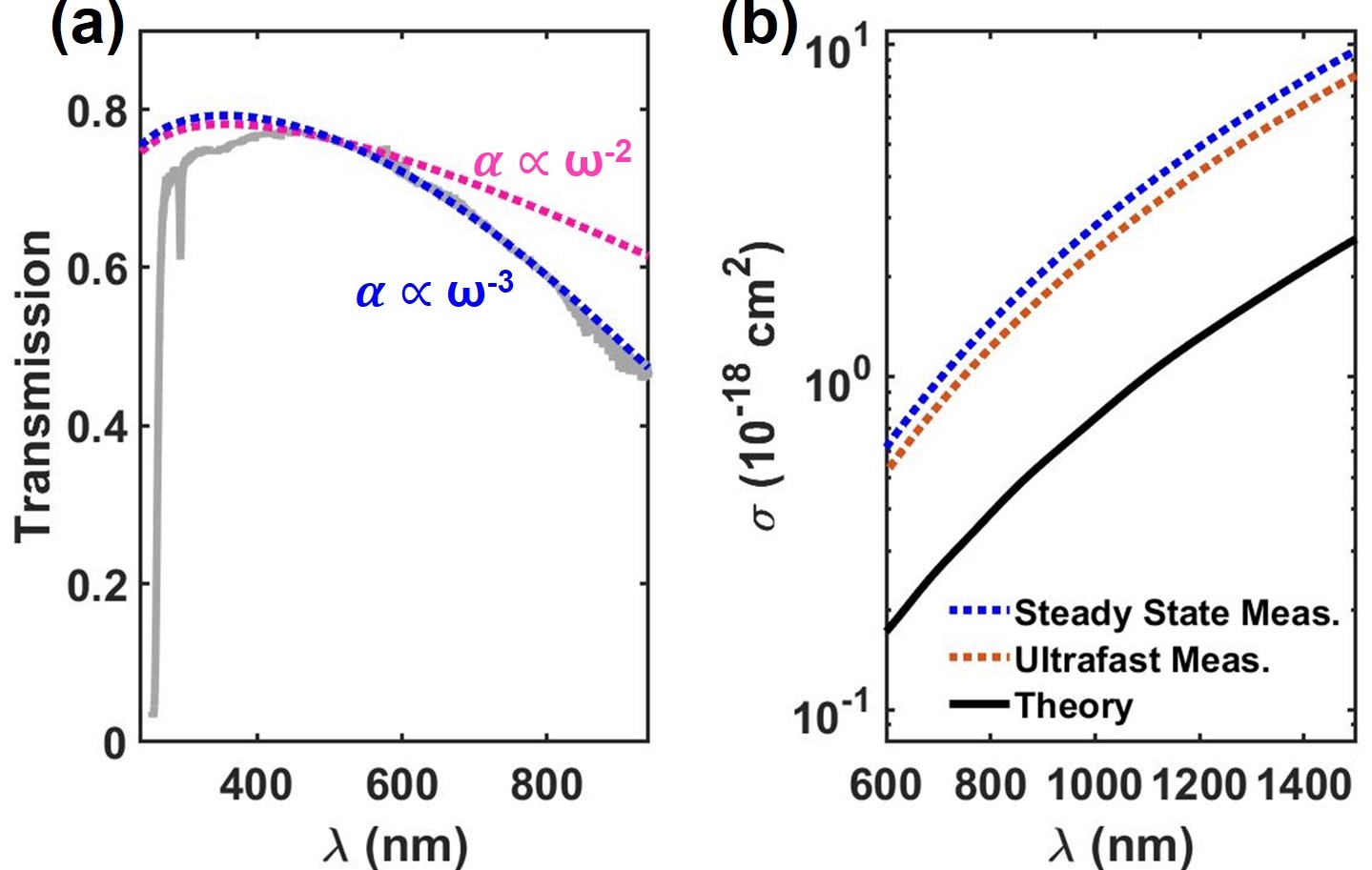}
		\caption{(a) Measured optical transmission spectrum (solid line) for 300 $\mu m$ thick, n-doped, ($\bar{2}01$) $\beta$-Ga$_2$O$_3$, for light polarized along the [010] axis, is plotted along with the fits obtained assuming inverse power law frequency dependence of the absorption coefficient. Transmission calculated assuming $\alpha \propto \omega^{-3}$ fits the data well over a broad wavelength range compared to that calculated assuming $\alpha \propto \omega^{-2}$ (Drude model). (b) Intra-band light absorption cross-sections (calculated by dividing the absorption coefficient with the free electron density) extracted from the steady state transmission measurements and ultrafast transmission measurements, and calculated using first principles \cite{peelaers_ga2o3fc}, are plotted as a function of wavelength.}
		\label{fig:trans_fit}
	\end{figure}

    We model light transmissivity $T$ through the sample with the simple equation, $T = e^{-{\alpha}L}(1-R)^2$, where $\alpha$ is the light absorption coefficient, $L$ is the sample thickness ($\sim 300$ $\mu$m), and $R$ is the reflectivity at the two facets of the sample. $R$ is calculated using the Fresnel equation assuming normal light incidence at the air - $\beta$-Ga$_2$O$_3$ interface. Wavelength-dependent values of the measured dielectric tensor\cite{sturm_dielectric_tensor} of $\beta$-Ga$_2$O$_3$ are used to calculate the reflectivity. We then fit the measured transmission spectra (from Figure \ref{fig:trans_pol_dep}) using absorption coefficients with an inverse power law frequency dependence (i.e. $\alpha \propto \omega^{-n}$). Figure \ref{fig:trans_fit}(a) shows the measured transmission spectrum for light polarized along the  [010] axis, along with fits to the data assuming $\omega^{-2}$ (Drude model) and  $\omega^{-3}$ frequency dependence of the absorption coefficient $\alpha$. As can be seen from this figure, an absorption coefficient $\alpha\propto\omega^{-3}$ fits the measured transmission spectrum well over a broad wavelength range. The observed frequency dependence of the absorption coefficient in our experiments confirms the frequency dependence predicted by theoretical calculations \cite{peelaers_ga2o3fc}.
    
    Intra-band optical absorption cross-section (calculated by dividing the absorption coefficient with the free electron density) extracted from the transmission measurements and calculated using first principles \cite{peelaers_ga2o3fc} are plotted as a function of wavelength in Figure \ref{fig:trans_fit}(b). Whereas the frequency dependence of the measured and calculated absorption cross-sections agree well, the theoretically calculated cross-section is smaller by a factor of $\sim 3$. This difference could be due to multiple reasons. Small differences in the curvature of the calculated bands might lead to an underestimated cross section. Further, the strength of the theoretically calculated electron-phonon coupling might be underestimated. Likely, a combination of the two factors is responsible for the difference.
    
    Despite the difference in the measured and calculated magnitudes, the agreement in wavelength dependence confirms that  longitudinal optical (LO) phonon emission is the mechanism behind the momentum conservation in the intra-band light absorption by free electrons in $\beta$-Ga$_2$O$_3$, in the visible to near-IR wavelength range \cite{peelaers_ga2o3fc}. This is not unexpected given the large LO phonon-electron coupling strength in $\beta$-Ga$_2$O$_3$ \cite{ma_mobility, Kang2017, ghosh_e-ph-coupling, mengle_e-ph-coupling}. The origin of the $1/\omega^{3}$ (rather than $1/\omega^{2}$) dependence of the absorption cross-section can be traced to the fact that far from the $\Gamma-$point in the $\beta$-Ga$_2$O$_3$ bandstructure (see Figure \ref{fig:bands}), the conduction band $E_{c}(k)$ becomes nearly {\em linear} in $k$ \cite{peelaers_ga2o3fc,peelaers_subbgabs}. Similar to our observations, this $1/\omega^{3}$ dependence has also been theoretically predicted and experimentally observed in SnO$_2$, another transparent conducting oxide \cite{peelaers_sno2fc, summitt_sno2fc}. Finally, the fact that this intra-band absorption strength does not depend on light polarization is expected given the near isotropy of the lowest conduction band electron states in $\beta$-Ga$_2$O$_3$ \cite{peelaers_bandstructure,Kang2017}.
        
    Impurities and defects can affect the measured wavelength dependence of the optical absorption. This is especially true for visible to near-IR absorption in wide bandgap materials. To isolate the optical absorption due to free carriers, it is thus preferable to measure free carrier absorption by modulating the conduction band electron density, and detecting only the resulting changes in the sub-bandgap light absorption. To accomplish this we measured time-resolved transmission through a photoexcited Sn-doped, 450 $\mu$m thick, bulk ($010$) $\beta$-Ga$_2$O$_3$ sample of free electron concentration $n \sim 5 \times10^{18}$ cm$^{-3}$ using ultrafast pump-probe spectroscopy. This sample was also grown using the EFG method by Tamura Corporation. We used a 405 nm center-wavelength ultrafast optical pulse as the pump that excited electrons from the valence band to the conduction band via a two-photon absorption process, as in our previous work \cite{koksal_opop}. The center-wavelength of the synchronized probe pulse, polarized along the crystal $a^{\star}$-axis (orthogonal to the $c$- and $b$-axes), was varied in the visible to the near-IR wavelength range. 
    
\begin{figure}[ht]
 		\includegraphics[width=1\columnwidth]{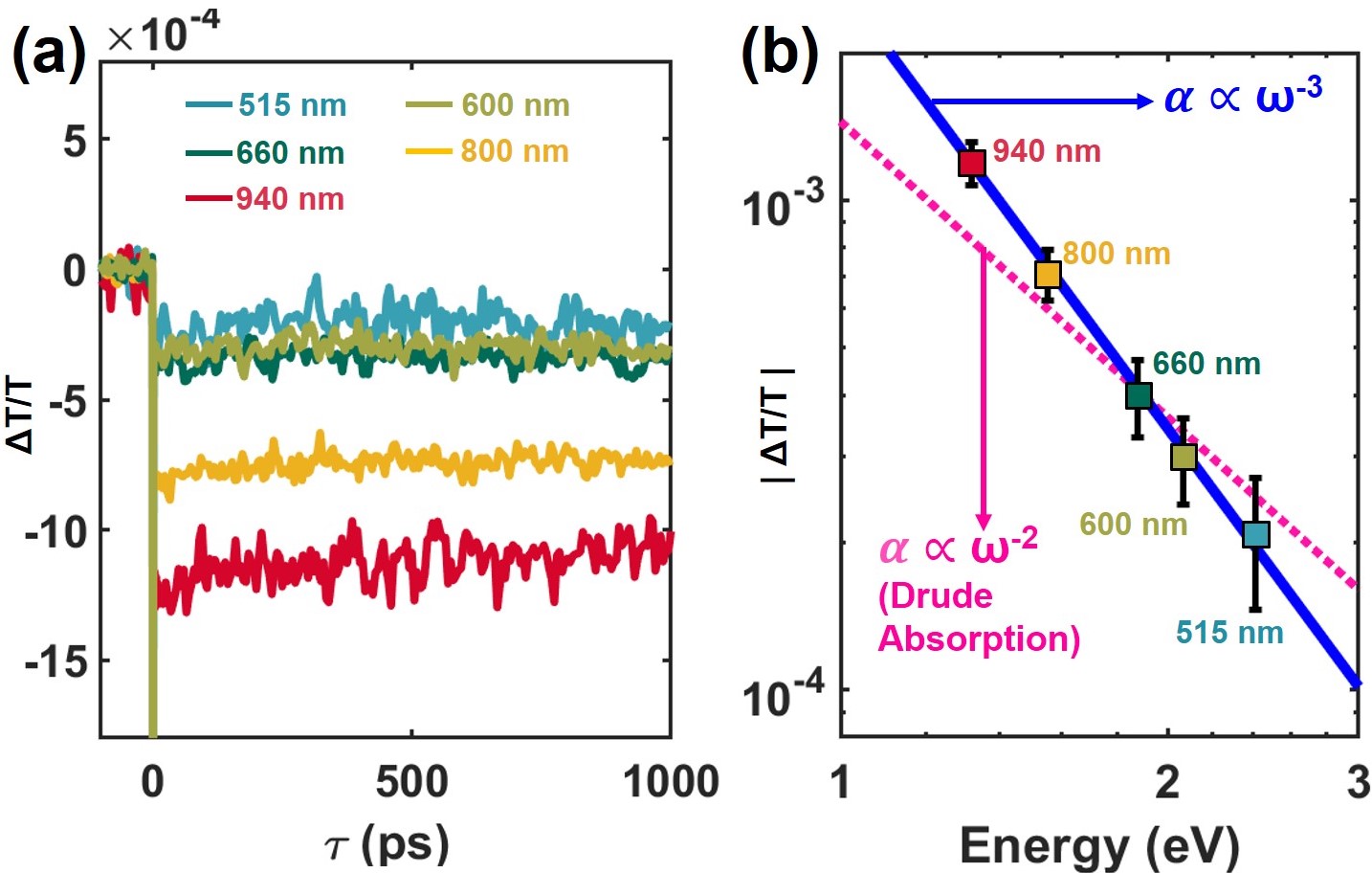}
 		\caption{(a) Differential probe pulse transmission ($\Delta T/T$) is plotted as a function of the probe delay (w.r.t. the 450 nm center-wavelength pump pulse) for different probe wavelengths. Results shown are for a Sn-doped, 450 $\mu m$ thick, bulk ($010$) $\beta$-Ga$_2$O$_3$ sample. The probe is always polarized along the crystal a*-axis. (b) $|\Delta T/T|$) for different probe wavelengths, for a fixed probe delay ($\sim 120$ ps), is plotted as a function of the probe photon energy. Also shown are the calculated best fit curves of $|\Delta T/T|$ assuming absorption coefficients with frequency dependence given by $\omega^{-3}$ and $\omega^{-2}$.}
 		\label{fig:opsp_spect}
 	\end{figure}

    Figure \ref{fig:opsp_spect}(a) shows the differential probe transmission ($\Delta T/T$) plotted as a function of the probe delay (w.r.t. the pump) for different probe wavelengths.  $\Delta T/T$ is negative because of optical absorption by photoexcited free electrons. The measured differential transmission magnitudes $|\Delta T/T|$ for different probe wavelengths, for a fixed probe delay ($\sim 120$ ps), are plotted as a function of the probe photon energy in Figure \ref{fig:opsp_spect}(b). Also plotted are the best fits for $|\Delta T/T|$, assuming absorption coefficients with frequency dependence given by $1/\omega^{3}$ and $1/\omega^{2}$. This figure shows that calculating $|\Delta T/T|$ using $\alpha\sim 1/\omega^{3}$ fits the data accurately. In addition, as shown in Figure \ref{fig:trans_fit}(b), the values of the absorption cross-section extracted from ultrafast measurements are in excellent agreement with those obtained using steady state transmission experiments.   
 	
    We now consider the inter-conduction band absorption features seen in Figure \ref{fig:trans_pol_dep}. This absorption is centered at $\lambda \sim 349$ nm (3.55 eV), and is polarization dependent. The largest absorption is seen for light polarized along the [102] axis. For light polarized along the orthogonal [010] axis, the inter-conduction band absorption is much weaker. As shown in Fig.~\ref{fig:bands} this absorption is due to transitions occurring between the lowest (partially filled) conduction band and the next conduction band, whose energy difference determines the optical absorption wavelength. The magnitude of the absorption depends on the dipole transition matrix elements, which are polarization dependent. According to the first-principles calculation of the bandstructure \cite{peelaers_subbgabs}, the inter-conduction band absorption for light polarized along the [010] axis should be much smaller than the absorption for light polarized along the [102] axis, which is in agreement with the experimental observation.
	
\begin{figure}[ht]
 		\includegraphics[width=1\columnwidth]{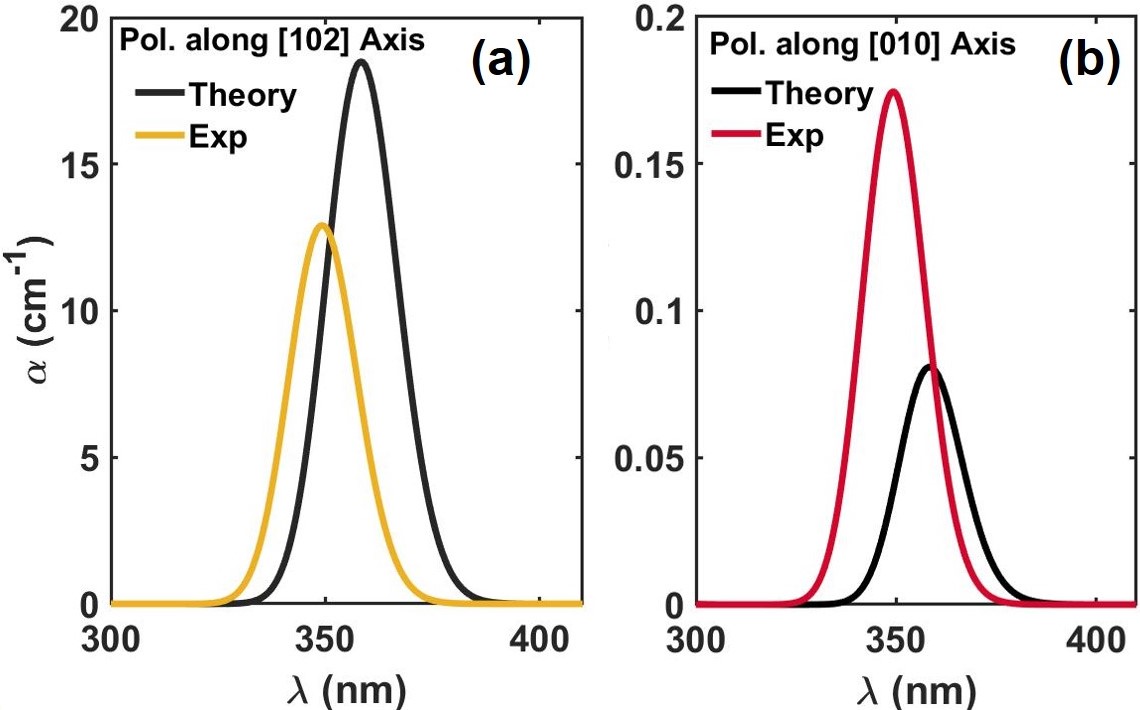}
 		\caption{The measured absorption coefficients for inter-conduction band absorption for light polarized along the [102] axis (a) and along the [010] axis (b) are plotted as a function of the wavelength. Also plotted are the absorption coefficients calculated from first principles \cite{peelaers_subbgabs}. The data is shown for a 300 $\mu$m thick, Sn-doped ($n\sim7\times10^{18}$ cm$^{-3}$), ($\bar{2}01$) $\beta$-Ga$_2$O$_3$ sample.}
 		\label{fig:icb_fit}
 	\end{figure}

	After subtracting a relatively small but broadband  absorption (partly due to intra-conduction band absorption), a Gaussian lineshape was used to fit the inter-conduction band absorption dips for light polarized along the [102] and [010] axes. Figure \ref{fig:icb_fit} shows the spectra of the inter-conduction band absorption coefficients extracted using this method, along with the theoretically calculated spectra from first principles \cite{peelaers_subbgabs}. For light polarization along the [102] and [010] axes, the experimental absorption coefficient values peak at $\sim 13$ cm$^{-1}$ and $\sim 0.17$ cm$^{-1}$, respectively. The agreement between experiments and theory is remarkable. The difference in the center wavelength of the absorption, between measurements and calculations, for both polarization directions is less than 10 nm (96 meV). This is well within the expected accuracy of the calculations. The difference in the peak absorption between measurement and theory is around 26\% for light polarization along the [102] axis, which is not unexpected given the experimental procedure used to extract the inter-conduction band absorption coefficient. The difference in the peak absorption between measurement and theory is larger for light polarization along the [010] axis; the measured absorption peak is almost twice as large as that predicted by theory. We should point out here that the peak absorption for light polarization along the [010] axis is extremely small (two orders of magnitude weaker than the absorption for light polarization along the [102] axis), and the observed dip in the transmission spectrum (Figure 2) is at the threshold of the signal to noise ratio available in our experimental setup. Therefore, a greater error would not be unexpected in quantifying the value of the absorption coefficient for light polarization along the [010] axis.

 	In conclusion, we have presented detailed experimental evidence of intra- and inter-conduction band optical absorption processes by free electrons in the conduction band of the ultra-wide bandgap semiconductor $\beta$-Ga$_2$O$_3$. Two unique features of this semiconductor were recently predicted by theoretical calculations: (1) sub-bandgap optical transitions in the UV from the bottom of the conduction band to the next conduction band with absorption coefficients in the $\sim 10$ cm$^{-1}$ range, and (2) an intra-conduction band transition that goes as $1/\omega^3$. Both these optical transitions are experimentally observed to be qualitatively and quantitatively consistent with the predictions. Since both the studied absorption mechanisms involve free electrons in the conduction band, a larger electron density would result in greater light absorption and reduced transparency in the UV, Visible, and Near-IR wavelength range. This is thus important to take into account in the design of optoelectronic devices relying on a high degree of transparency. Further, electrons in the lowest conduction band states can scatter into the higher conduction band under high electric fields, and the absorption measurements provide a measure of the matrix elements for such transitions.

 	\begin{acknowledgments}
 	*Singh and Koksal contributed equally to this work. This work was supported by NSF under Grant No. DMR-1719875 and AFOSR under Grant No. FA9550-17-1-0048. 
 	\end{acknowledgments}
 
 	\section*{Data Availability Statement}
 	The data that support the findings of this study are available from the corresponding author upon reasonable request.

	\bibliography{ga2o3_fc_citations}

\begin{thebibliography}{32}%
\makeatletter
\providecommand \@ifxundefined [1]{%
 \@ifx{#1\undefined}
}%
\providecommand \@ifnum [1]{%
 \ifnum #1\expandafter \@firstoftwo
 \else \expandafter \@secondoftwo
 \fi
}%
\providecommand \@ifx [1]{%
 \ifx #1\expandafter \@firstoftwo
 \else \expandafter \@secondoftwo
 \fi
}%
\providecommand \natexlab [1]{#1}%
\providecommand \enquote  [1]{``#1''}%
\providecommand \bibnamefont  [1]{#1}%
\providecommand \bibfnamefont [1]{#1}%
\providecommand \citenamefont [1]{#1}%
\providecommand \href@noop [0]{\@secondoftwo}%
\providecommand \href [0]{\begingroup \@sanitize@url \@href}%
\providecommand \@href[1]{\@@startlink{#1}\@@href}%
\providecommand \@@href[1]{\endgroup#1\@@endlink}%
\providecommand \@sanitize@url [0]{\catcode `\\12\catcode `\$12\catcode
  `\&12\catcode `\#12\catcode `\^12\catcode `\_12\catcode `\%12\relax}%
\providecommand \@@startlink[1]{}%
\providecommand \@@endlink[0]{}%
\providecommand \url  [0]{\begingroup\@sanitize@url \@url }%
\providecommand \@url [1]{\endgroup\@href {#1}{\urlprefix }}%
\providecommand \urlprefix  [0]{URL }%
\providecommand \Eprint [0]{\href }%
\providecommand \doibase [0]{http://dx.doi.org/}%
\providecommand \selectlanguage [0]{\@gobble}%
\providecommand \bibinfo  [0]{\@secondoftwo}%
\providecommand \bibfield  [0]{\@secondoftwo}%
\providecommand \translation [1]{[#1]}%
\providecommand \BibitemOpen [0]{}%
\providecommand \bibitemStop [0]{}%
\providecommand \bibitemNoStop [0]{.\EOS\space}%
\providecommand \EOS [0]{\spacefactor3000\relax}%
\providecommand \BibitemShut  [1]{\csname bibitem#1\endcsname}%
\let\auto@bib@innerbib\@empty
\bibitem [{\citenamefont {Ricci}\ \emph {et~al.}(2016)\citenamefont {Ricci},
  \citenamefont {Boschi}, \citenamefont {Baraldi}, \citenamefont {Filippetti},
  \citenamefont {Higashiwaki}, \citenamefont {Kuramata}, \citenamefont
  {Fiorentini},\ and\ \citenamefont {Fornari}}]{ricci_bg}%
  \BibitemOpen
  \bibfield  {author} {\bibinfo {author} {\bibfnamefont {F.}~\bibnamefont
  {Ricci}}, \bibinfo {author} {\bibfnamefont {F.}~\bibnamefont {Boschi}},
  \bibinfo {author} {\bibfnamefont {A.}~\bibnamefont {Baraldi}}, \bibinfo
  {author} {\bibfnamefont {A.}~\bibnamefont {Filippetti}}, \bibinfo {author}
  {\bibfnamefont {M.}~\bibnamefont {Higashiwaki}}, \bibinfo {author}
  {\bibfnamefont {A.}~\bibnamefont {Kuramata}}, \bibinfo {author}
  {\bibfnamefont {V.}~\bibnamefont {Fiorentini}}, \ and\ \bibinfo {author}
  {\bibfnamefont {R.}~\bibnamefont {Fornari}},\ }\href@noop {} {\bibfield
  {journal} {\bibinfo  {journal} {Journal of Physics: Condensed Matter}\
  }\textbf {\bibinfo {volume} {28}},\ \bibinfo {pages} {224005} (\bibinfo
  {year} {2016})}\BibitemShut {NoStop}%
\bibitem [{\citenamefont {Orita}\ \emph {et~al.}(2000)\citenamefont {Orita},
  \citenamefont {Ohta}, \citenamefont {Hirano},\ and\ \citenamefont
  {Hosono}}]{masahiro_bg}%
  \BibitemOpen
  \bibfield  {author} {\bibinfo {author} {\bibfnamefont {M.}~\bibnamefont
  {Orita}}, \bibinfo {author} {\bibfnamefont {H.}~\bibnamefont {Ohta}},
  \bibinfo {author} {\bibfnamefont {M.}~\bibnamefont {Hirano}}, \ and\ \bibinfo
  {author} {\bibfnamefont {H.}~\bibnamefont {Hosono}},\ }\href@noop {}
  {\bibfield  {journal} {\bibinfo  {journal} {Applied Physics Letters}\
  }\textbf {\bibinfo {volume} {77}},\ \bibinfo {pages} {4166--4168} (\bibinfo
  {year} {2000})}\BibitemShut {NoStop}%
\bibitem [{\citenamefont {Matsumoto}\ \emph {et~al.}(1974)\citenamefont
  {Matsumoto}, \citenamefont {Aoki}, \citenamefont {Kinoshita},\ and\
  \citenamefont {Aono}}]{matsumoto_bg}%
  \BibitemOpen
  \bibfield  {author} {\bibinfo {author} {\bibfnamefont {T.}~\bibnamefont
  {Matsumoto}}, \bibinfo {author} {\bibfnamefont {M.}~\bibnamefont {Aoki}},
  \bibinfo {author} {\bibfnamefont {A.}~\bibnamefont {Kinoshita}}, \ and\
  \bibinfo {author} {\bibfnamefont {T.}~\bibnamefont {Aono}},\ }\href@noop {}
  {\bibfield  {journal} {\bibinfo  {journal} {Japanese Journal of Applied
  Physics}\ }\textbf {\bibinfo {volume} {13}},\ \bibinfo {pages} {1578--1582}
  (\bibinfo {year} {1974})}\BibitemShut {NoStop}%
\bibitem [{\citenamefont {Tippins}(1965)}]{tippins_bg}%
  \BibitemOpen
  \bibfield  {author} {\bibinfo {author} {\bibfnamefont {H.~H.}\ \bibnamefont
  {Tippins}},\ }\href@noop {} {\bibfield  {journal} {\bibinfo  {journal} {Phys.
  Rev.}\ }\textbf {\bibinfo {volume} {140}},\ \bibinfo {pages} {A316--A319}
  (\bibinfo {year} {1965})}\BibitemShut {NoStop}%
\bibitem [{\citenamefont {Aida}\ \emph {et~al.}(2008)\citenamefont {Aida},
  \citenamefont {Nishiguchi}, \citenamefont {Takeda}, \citenamefont {Aota},
  \citenamefont {Sunakawa},\ and\ \citenamefont {Yaguchi}}]{aida_efg}%
  \BibitemOpen
  \bibfield  {author} {\bibinfo {author} {\bibfnamefont {H.}~\bibnamefont
  {Aida}}, \bibinfo {author} {\bibfnamefont {K.}~\bibnamefont {Nishiguchi}},
  \bibinfo {author} {\bibfnamefont {H.}~\bibnamefont {Takeda}}, \bibinfo
  {author} {\bibfnamefont {N.}~\bibnamefont {Aota}}, \bibinfo {author}
  {\bibfnamefont {K.}~\bibnamefont {Sunakawa}}, \ and\ \bibinfo {author}
  {\bibfnamefont {Y.}~\bibnamefont {Yaguchi}},\ }\href@noop {} {\bibfield
  {journal} {\bibinfo  {journal} {Japanese Journal of Applied Physics}\
  }\textbf {\bibinfo {volume} {47}},\ \bibinfo {pages} {8506--8509} (\bibinfo
  {year} {2008})}\BibitemShut {NoStop}%
\bibitem [{\citenamefont {Kuramata}\ \emph {et~al.}(2016)\citenamefont
  {Kuramata}, \citenamefont {Koshi}, \citenamefont {Watanabe}, \citenamefont
  {Yamaoka}, \citenamefont {Masui},\ and\ \citenamefont
  {Yamakoshi}}]{kuramata_efg}%
  \BibitemOpen
  \bibfield  {author} {\bibinfo {author} {\bibfnamefont {A.}~\bibnamefont
  {Kuramata}}, \bibinfo {author} {\bibfnamefont {K.}~\bibnamefont {Koshi}},
  \bibinfo {author} {\bibfnamefont {S.}~\bibnamefont {Watanabe}}, \bibinfo
  {author} {\bibfnamefont {Y.}~\bibnamefont {Yamaoka}}, \bibinfo {author}
  {\bibfnamefont {T.}~\bibnamefont {Masui}}, \ and\ \bibinfo {author}
  {\bibfnamefont {S.}~\bibnamefont {Yamakoshi}},\ }\href@noop {} {\bibfield
  {journal} {\bibinfo  {journal} {Japanese Journal of Applied Physics}\
  }\textbf {\bibinfo {volume} {55}},\ \bibinfo {pages} {1202A2} (\bibinfo
  {year} {2016})}\BibitemShut {NoStop}%
\bibitem [{\citenamefont {Galazka}\ \emph {et~al.}(2014)\citenamefont
  {Galazka}, \citenamefont {Irmscher}, \citenamefont {Uecker}, \citenamefont
  {Bertram}, \citenamefont {Pietsch}, \citenamefont {Kwasniewski},
  \citenamefont {Naumann}, \citenamefont {Schulz}, \citenamefont {Schewski},
  \citenamefont {Klimm},\ and\ \citenamefont
  {Bickermann}}]{galazga_czochralski}%
  \BibitemOpen
  \bibfield  {author} {\bibinfo {author} {\bibfnamefont {Z.}~\bibnamefont
  {Galazka}}, \bibinfo {author} {\bibfnamefont {K.}~\bibnamefont {Irmscher}},
  \bibinfo {author} {\bibfnamefont {R.}~\bibnamefont {Uecker}}, \bibinfo
  {author} {\bibfnamefont {R.}~\bibnamefont {Bertram}}, \bibinfo {author}
  {\bibfnamefont {M.}~\bibnamefont {Pietsch}}, \bibinfo {author} {\bibfnamefont
  {A.}~\bibnamefont {Kwasniewski}}, \bibinfo {author} {\bibfnamefont
  {M.}~\bibnamefont {Naumann}}, \bibinfo {author} {\bibfnamefont
  {T.}~\bibnamefont {Schulz}}, \bibinfo {author} {\bibfnamefont
  {R.}~\bibnamefont {Schewski}}, \bibinfo {author} {\bibfnamefont
  {D.}~\bibnamefont {Klimm}}, \ and\ \bibinfo {author} {\bibfnamefont
  {M.}~\bibnamefont {Bickermann}},\ }\href@noop {} {\bibfield  {journal}
  {\bibinfo  {journal} {Journal of Crystal Growth}\ }\textbf {\bibinfo {volume}
  {404}},\ \bibinfo {pages} {184--191} (\bibinfo {year} {2014})}\BibitemShut
  {NoStop}%
\bibitem [{\citenamefont {Hu}\ \emph {et~al.}(2015)\citenamefont {Hu},
  \citenamefont {Shan}, \citenamefont {Zhang}, \citenamefont {Jiang},
  \citenamefont {Wang},\ and\ \citenamefont {Shen}}]{hu_uvdet}%
  \BibitemOpen
  \bibfield  {author} {\bibinfo {author} {\bibfnamefont {G.~C.}\ \bibnamefont
  {Hu}}, \bibinfo {author} {\bibfnamefont {C.~X.}\ \bibnamefont {Shan}},
  \bibinfo {author} {\bibfnamefont {N.}~\bibnamefont {Zhang}}, \bibinfo
  {author} {\bibfnamefont {M.~M.}\ \bibnamefont {Jiang}}, \bibinfo {author}
  {\bibfnamefont {S.~P.}\ \bibnamefont {Wang}}, \ and\ \bibinfo {author}
  {\bibfnamefont {D.~Z.}\ \bibnamefont {Shen}},\ }\href@noop {} {\bibfield
  {journal} {\bibinfo  {journal} {Opt. Express}\ }\textbf {\bibinfo {volume}
  {23}},\ \bibinfo {pages} {13554--13561} (\bibinfo {year} {2015})}\BibitemShut
  {NoStop}%
\bibitem [{\citenamefont {Ji}\ \emph {et~al.}(2006)\citenamefont {Ji},
  \citenamefont {Du}, \citenamefont {Fan},\ and\ \citenamefont
  {Wang}}]{ji_uvdet}%
  \BibitemOpen
  \bibfield  {author} {\bibinfo {author} {\bibfnamefont {Z.}~\bibnamefont
  {Ji}}, \bibinfo {author} {\bibfnamefont {J.}~\bibnamefont {Du}}, \bibinfo
  {author} {\bibfnamefont {J.}~\bibnamefont {Fan}}, \ and\ \bibinfo {author}
  {\bibfnamefont {W.}~\bibnamefont {Wang}},\ }\href@noop {} {\bibfield
  {journal} {\bibinfo  {journal} {Optical Materials}\ }\textbf {\bibinfo
  {volume} {28}},\ \bibinfo {pages} {415 -- 417} (\bibinfo {year}
  {2006})}\BibitemShut {NoStop}%
\bibitem [{\citenamefont {Oshima}, \citenamefont {Okuno},\ and\ \citenamefont
  {Fujita}(2007)}]{oshima_uvdet}%
  \BibitemOpen
  \bibfield  {author} {\bibinfo {author} {\bibfnamefont {T.}~\bibnamefont
  {Oshima}}, \bibinfo {author} {\bibfnamefont {T.}~\bibnamefont {Okuno}}, \
  and\ \bibinfo {author} {\bibfnamefont {S.}~\bibnamefont {Fujita}},\
  }\href@noop {} {\bibfield  {journal} {\bibinfo  {journal} {Japanese Journal
  of Applied Physics}\ }\textbf {\bibinfo {volume} {46}},\ \bibinfo {pages}
  {7217--7220} (\bibinfo {year} {2007})}\BibitemShut {NoStop}%
\bibitem [{\citenamefont {Minami}, \citenamefont {Nishi},\ and\ \citenamefont
  {Miyata}(2013)}]{minami_solarcell}%
  \BibitemOpen
  \bibfield  {author} {\bibinfo {author} {\bibfnamefont {T.}~\bibnamefont
  {Minami}}, \bibinfo {author} {\bibfnamefont {Y.}~\bibnamefont {Nishi}}, \
  and\ \bibinfo {author} {\bibfnamefont {T.}~\bibnamefont {Miyata}},\ }\href
  {\doibase 10.7567/apex.6.044101} {\bibfield  {journal} {\bibinfo  {journal}
  {Applied Physics Express}\ }\textbf {\bibinfo {volume} {6}},\ \bibinfo
  {pages} {044101} (\bibinfo {year} {2013})}\BibitemShut {NoStop}%
\bibitem [{\citenamefont {Haga}\ and\ \citenamefont
  {Kimura}(1964)}]{eijiro_intercb_III_V}%
  \BibitemOpen
  \bibfield  {author} {\bibinfo {author} {\bibfnamefont {E.}~\bibnamefont
  {Haga}}\ and\ \bibinfo {author} {\bibfnamefont {H.}~\bibnamefont {Kimura}},\
  }\href@noop {} {\bibfield  {journal} {\bibinfo  {journal} {Journal of the
  Physical Society of Japan}\ }\textbf {\bibinfo {volume} {19}},\ \bibinfo
  {pages} {1596--1606} (\bibinfo {year} {1964})}\BibitemShut {NoStop}%
\bibitem [{\citenamefont {Glosser}, \citenamefont {Fischer},\ and\
  \citenamefont {Seraphin}(1970)}]{glosser_intercb_insb}%
  \BibitemOpen
  \bibfield  {author} {\bibinfo {author} {\bibfnamefont {R.}~\bibnamefont
  {Glosser}}, \bibinfo {author} {\bibfnamefont {J.~E.}\ \bibnamefont
  {Fischer}}, \ and\ \bibinfo {author} {\bibfnamefont {B.~O.}\ \bibnamefont
  {Seraphin}},\ }\href@noop {} {\bibfield  {journal} {\bibinfo  {journal}
  {Phys. Rev. B}\ }\textbf {\bibinfo {volume} {1}},\ \bibinfo {pages}
  {1607--1610} (\bibinfo {year} {1970})}\BibitemShut {NoStop}%
\bibitem [{\citenamefont {Balkanski}, \citenamefont {Aziza},\ and\
  \citenamefont {Amzallag}(1969)}]{balkanski_intercb_silicon}%
  \BibitemOpen
  \bibfield  {author} {\bibinfo {author} {\bibfnamefont {M.}~\bibnamefont
  {Balkanski}}, \bibinfo {author} {\bibfnamefont {A.}~\bibnamefont {Aziza}}, \
  and\ \bibinfo {author} {\bibfnamefont {E.}~\bibnamefont {Amzallag}},\
  }\href@noop {} {\bibfield  {journal} {\bibinfo  {journal} {physica status
  solidi (b)}\ }\textbf {\bibinfo {volume} {31}},\ \bibinfo {pages} {323--330}
  (\bibinfo {year} {1969})}\BibitemShut {NoStop}%
\bibitem [{\citenamefont {Canestraro}\ \emph {et~al.}(2008)\citenamefont
  {Canestraro}, \citenamefont {Oliveira}, \citenamefont {Valaski},
  \citenamefont {{da Silva}}, \citenamefont {David}, \citenamefont {Pepe},
  \citenamefont {da~Silva}, \citenamefont {Roman},\ and\ \citenamefont
  {Persson}}]{canestraro_intercb_FTO}%
  \BibitemOpen
  \bibfield  {author} {\bibinfo {author} {\bibfnamefont {C.~D.}\ \bibnamefont
  {Canestraro}}, \bibinfo {author} {\bibfnamefont {M.~M.}\ \bibnamefont
  {Oliveira}}, \bibinfo {author} {\bibfnamefont {R.}~\bibnamefont {Valaski}},
  \bibinfo {author} {\bibfnamefont {M.~V.}\ \bibnamefont {{da Silva}}},
  \bibinfo {author} {\bibfnamefont {D.~G.}\ \bibnamefont {David}}, \bibinfo
  {author} {\bibfnamefont {I.}~\bibnamefont {Pepe}}, \bibinfo {author}
  {\bibfnamefont {A.~F.}\ \bibnamefont {da~Silva}}, \bibinfo {author}
  {\bibfnamefont {L.~S.}\ \bibnamefont {Roman}}, \ and\ \bibinfo {author}
  {\bibfnamefont {C.}~\bibnamefont {Persson}},\ }\href@noop {} {\bibfield
  {journal} {\bibinfo  {journal} {Applied Surface Science}\ }\textbf {\bibinfo
  {volume} {255}},\ \bibinfo {pages} {1874 -- 1879} (\bibinfo {year}
  {2008})}\BibitemShut {NoStop}%
\bibitem [{\citenamefont {Baker-Finch}\ \emph {et~al.}(2014)\citenamefont
  {Baker-Finch}, \citenamefont {McIntosh}, \citenamefont {Yan}, \citenamefont
  {Fong},\ and\ \citenamefont {Kho}}]{baker-finch_sifc}%
  \BibitemOpen
  \bibfield  {author} {\bibinfo {author} {\bibfnamefont {S.~C.}\ \bibnamefont
  {Baker-Finch}}, \bibinfo {author} {\bibfnamefont {K.~R.}\ \bibnamefont
  {McIntosh}}, \bibinfo {author} {\bibfnamefont {D.}~\bibnamefont {Yan}},
  \bibinfo {author} {\bibfnamefont {K.~C.}\ \bibnamefont {Fong}}, \ and\
  \bibinfo {author} {\bibfnamefont {T.~C.}\ \bibnamefont {Kho}},\ }\href
  {\doibase 10.1063/1.4893176} {\bibfield  {journal} {\bibinfo  {journal}
  {Journal of Applied Physics}\ }\textbf {\bibinfo {volume} {116}},\ \bibinfo
  {pages} {063106} (\bibinfo {year} {2014})}\BibitemShut {NoStop}%
\bibitem [{\citenamefont {Perkowitz}(1969)}]{perkowitz_gaasfc1}%
  \BibitemOpen
  \bibfield  {author} {\bibinfo {author} {\bibfnamefont {S.}~\bibnamefont
  {Perkowitz}},\ }\href@noop {} {\bibfield  {journal} {\bibinfo  {journal}
  {Journal of Applied Physics}\ }\textbf {\bibinfo {volume} {40}},\ \bibinfo
  {pages} {3751--3754} (\bibinfo {year} {1969})}\BibitemShut {NoStop}%
\bibitem [{\citenamefont {Perkowitz}(1971)}]{perkowitz_gaasfc2}%
  \BibitemOpen
  \bibfield  {author} {\bibinfo {author} {\bibfnamefont {S.}~\bibnamefont
  {Perkowitz}},\ }\href@noop {} {\bibfield  {journal} {\bibinfo  {journal}
  {Journal of Physics and Chemistry of Solids}\ }\textbf {\bibinfo {volume}
  {32}},\ \bibinfo {pages} {2267 -- 2274} (\bibinfo {year} {1971})}\BibitemShut
  {NoStop}%
\bibitem [{\citenamefont {von Baltz}\ and\ \citenamefont
  {Escher}(1972)}]{von-baltz_qmfc}%
  \BibitemOpen
  \bibfield  {author} {\bibinfo {author} {\bibfnamefont {R.}~\bibnamefont {von
  Baltz}}\ and\ \bibinfo {author} {\bibfnamefont {W.}~\bibnamefont {Escher}},\
  }\href@noop {} {\bibfield  {journal} {\bibinfo  {journal} {physica status
  solidi (b)}\ }\textbf {\bibinfo {volume} {51}},\ \bibinfo {pages} {499--507}
  (\bibinfo {year} {1972})}\BibitemShut {NoStop}%
\bibitem [{\citenamefont {Peelaers}, \citenamefont {Kioupakis},\ and\
  \citenamefont {Van~de Walle}(2012)}]{peelaers_sno2fc}%
  \BibitemOpen
  \bibfield  {author} {\bibinfo {author} {\bibfnamefont {H.}~\bibnamefont
  {Peelaers}}, \bibinfo {author} {\bibfnamefont {E.}~\bibnamefont {Kioupakis}},
  \ and\ \bibinfo {author} {\bibfnamefont {C.~G.}\ \bibnamefont {Van~de
  Walle}},\ }\href@noop {} {\bibfield  {journal} {\bibinfo  {journal} {Applied
  Physics Letters}\ }\textbf {\bibinfo {volume} {100}},\ \bibinfo {pages}
  {011914} (\bibinfo {year} {2012})}\BibitemShut {NoStop}%
\bibitem [{\citenamefont {Peelaers}\ and\ \citenamefont {Van~de
  Walle}(2019)}]{peelaers_ga2o3fc}%
  \BibitemOpen
  \bibfield  {author} {\bibinfo {author} {\bibfnamefont {H.}~\bibnamefont
  {Peelaers}}\ and\ \bibinfo {author} {\bibfnamefont {C.~G.}\ \bibnamefont
  {Van~de Walle}},\ }\href@noop {} {\bibfield  {journal} {\bibinfo  {journal}
  {Phys. Rev. B}\ }\textbf {\bibinfo {volume} {100}},\ \bibinfo {pages}
  {081202} (\bibinfo {year} {2019})}\BibitemShut {NoStop}%
\bibitem [{\citenamefont {Peelaers}\ and\ \citenamefont {Van~de
  Walle}(2017)}]{peelaers_subbgabs}%
  \BibitemOpen
  \bibfield  {author} {\bibinfo {author} {\bibfnamefont {H.}~\bibnamefont
  {Peelaers}}\ and\ \bibinfo {author} {\bibfnamefont {C.~G.}\ \bibnamefont
  {Van~de Walle}},\ }\href {\doibase 10.1063/1.5001323} {\bibfield  {journal}
  {\bibinfo  {journal} {Applied Physics Letters}\ }\textbf {\bibinfo {volume}
  {111}},\ \bibinfo {pages} {182104} (\bibinfo {year} {2017})}\BibitemShut
  {NoStop}%
\bibitem [{\citenamefont {Dong}\ \emph {et~al.}(2017)\citenamefont {Dong},
  \citenamefont {Jia}, \citenamefont {Xin}, \citenamefont {Peng},\ and\
  \citenamefont {Zhang}}]{dong_optical_defects}%
  \BibitemOpen
  \bibfield  {author} {\bibinfo {author} {\bibfnamefont {L.}~\bibnamefont
  {Dong}}, \bibinfo {author} {\bibfnamefont {R.}~\bibnamefont {Jia}}, \bibinfo
  {author} {\bibfnamefont {B.}~\bibnamefont {Xin}}, \bibinfo {author}
  {\bibfnamefont {B.}~\bibnamefont {Peng}}, \ and\ \bibinfo {author}
  {\bibfnamefont {Y.}~\bibnamefont {Zhang}},\ }\href@noop {} {\bibfield
  {journal} {\bibinfo  {journal} {Scientific Reports}\ }\textbf {\bibinfo
  {volume} {7}},\ \bibinfo {pages} {40160} (\bibinfo {year}
  {2017})}\BibitemShut {NoStop}%
\bibitem [{\citenamefont {Peelaers}, \citenamefont {Kioupakis},\ and\
  \citenamefont {{Van de Walle}}(2019)}]{Peelaers2019b}%
  \BibitemOpen
  \bibfield  {author} {\bibinfo {author} {\bibfnamefont {H.}~\bibnamefont
  {Peelaers}}, \bibinfo {author} {\bibfnamefont {E.}~\bibnamefont {Kioupakis}},
  \ and\ \bibinfo {author} {\bibfnamefont {C.~G.}\ \bibnamefont {{Van de
  Walle}}},\ }\href {\doibase 10.1063/1.5109569} {\bibfield  {journal}
  {\bibinfo  {journal} {Applied Physics Letters}\ }\textbf {\bibinfo {volume}
  {115}},\ \bibinfo {pages} {082105} (\bibinfo {year} {2019})}\BibitemShut
  {NoStop}%
\bibitem [{\citenamefont {Sturm}\ \emph {et~al.}(2015)\citenamefont {Sturm},
  \citenamefont {Furthmüller}, \citenamefont {Bechstedt}, \citenamefont
  {Schmidt-Grund},\ and\ \citenamefont {Grundmann}}]{sturm_dielectric_tensor}%
  \BibitemOpen
  \bibfield  {author} {\bibinfo {author} {\bibfnamefont {C.}~\bibnamefont
  {Sturm}}, \bibinfo {author} {\bibfnamefont {J.}~\bibnamefont {Furthmüller}},
  \bibinfo {author} {\bibfnamefont {F.}~\bibnamefont {Bechstedt}}, \bibinfo
  {author} {\bibfnamefont {R.}~\bibnamefont {Schmidt-Grund}}, \ and\ \bibinfo
  {author} {\bibfnamefont {M.}~\bibnamefont {Grundmann}},\ }\href@noop {}
  {\bibfield  {journal} {\bibinfo  {journal} {APL Materials}\ }\textbf
  {\bibinfo {volume} {3}},\ \bibinfo {pages} {106106} (\bibinfo {year}
  {2015})}\BibitemShut {NoStop}%
\bibitem [{\citenamefont {Ma}\ \emph {et~al.}(2016)\citenamefont {Ma},
  \citenamefont {Tanen}, \citenamefont {Verma}, \citenamefont {Guo},
  \citenamefont {Luo}, \citenamefont {Xing},\ and\ \citenamefont
  {Jena}}]{ma_mobility}%
  \BibitemOpen
  \bibfield  {author} {\bibinfo {author} {\bibfnamefont {N.}~\bibnamefont
  {Ma}}, \bibinfo {author} {\bibfnamefont {N.}~\bibnamefont {Tanen}}, \bibinfo
  {author} {\bibfnamefont {A.}~\bibnamefont {Verma}}, \bibinfo {author}
  {\bibfnamefont {Z.}~\bibnamefont {Guo}}, \bibinfo {author} {\bibfnamefont
  {T.}~\bibnamefont {Luo}}, \bibinfo {author} {\bibfnamefont {H.~G.}\
  \bibnamefont {Xing}}, \ and\ \bibinfo {author} {\bibfnamefont
  {D.}~\bibnamefont {Jena}},\ }\href@noop {} {\bibfield  {journal} {\bibinfo
  {journal} {Applied Physics Letters}\ }\textbf {\bibinfo {volume} {109}},\
  \bibinfo {pages} {212101} (\bibinfo {year} {2016})}\BibitemShut {NoStop}%
\bibitem [{\citenamefont {Kang}\ \emph {et~al.}(2017)\citenamefont {Kang},
  \citenamefont {Krishnaswamy}, \citenamefont {Peelaers},\ and\ \citenamefont
  {{Van de Walle}}}]{Kang2017}%
  \BibitemOpen
  \bibfield  {author} {\bibinfo {author} {\bibfnamefont {Y.}~\bibnamefont
  {Kang}}, \bibinfo {author} {\bibfnamefont {K.}~\bibnamefont {Krishnaswamy}},
  \bibinfo {author} {\bibfnamefont {H.}~\bibnamefont {Peelaers}}, \ and\
  \bibinfo {author} {\bibfnamefont {C.~G.}\ \bibnamefont {{Van de Walle}}},\
  }\href@noop {} {\bibfield  {journal} {\bibinfo  {journal} {Journal of
  Physics: Condensed Matter}\ }\textbf {\bibinfo {volume} {29}},\ \bibinfo
  {pages} {234001--234001} (\bibinfo {year} {2017})}\BibitemShut {NoStop}%
\bibitem [{\citenamefont {Ghosh}\ and\ \citenamefont
  {Singisetti}(2016)}]{ghosh_e-ph-coupling}%
  \BibitemOpen
  \bibfield  {author} {\bibinfo {author} {\bibfnamefont {K.}~\bibnamefont
  {Ghosh}}\ and\ \bibinfo {author} {\bibfnamefont {U.}~\bibnamefont
  {Singisetti}},\ }\href@noop {} {\bibfield  {journal} {\bibinfo  {journal}
  {Applied Physics Letters}\ }\textbf {\bibinfo {volume} {109}},\ \bibinfo
  {pages} {072102} (\bibinfo {year} {2016})}\BibitemShut {NoStop}%
\bibitem [{\citenamefont {Mengle}\ and\ \citenamefont
  {Kioupakis}(2019)}]{mengle_e-ph-coupling}%
  \BibitemOpen
  \bibfield  {author} {\bibinfo {author} {\bibfnamefont {K.~A.}\ \bibnamefont
  {Mengle}}\ and\ \bibinfo {author} {\bibfnamefont {E.}~\bibnamefont
  {Kioupakis}},\ }\href@noop {} {\bibfield  {journal} {\bibinfo  {journal} {AIP
  Advances}\ }\textbf {\bibinfo {volume} {9}},\ \bibinfo {pages} {015313}
  (\bibinfo {year} {2019})}\BibitemShut {NoStop}%
\bibitem [{\citenamefont {Summitt}\ and\ \citenamefont
  {Borrelli}(1965)}]{summitt_sno2fc}%
  \BibitemOpen
  \bibfield  {author} {\bibinfo {author} {\bibfnamefont {R.}~\bibnamefont
  {Summitt}}\ and\ \bibinfo {author} {\bibfnamefont {N.}~\bibnamefont
  {Borrelli}},\ }\href@noop {} {\bibfield  {journal} {\bibinfo  {journal}
  {Journal of Physics and Chemistry of Solids}\ }\textbf {\bibinfo {volume}
  {26}},\ \bibinfo {pages} {921 -- 925} (\bibinfo {year} {1965})}\BibitemShut
  {NoStop}%
\bibitem [{\citenamefont {Peelaers}\ and\ \citenamefont {Van~de
  Walle}(2015)}]{peelaers_bandstructure}%
  \BibitemOpen
  \bibfield  {author} {\bibinfo {author} {\bibfnamefont {H.}~\bibnamefont
  {Peelaers}}\ and\ \bibinfo {author} {\bibfnamefont {C.~G.}\ \bibnamefont
  {Van~de Walle}},\ }\href@noop {} {\bibfield  {journal} {\bibinfo  {journal}
  {physica status solidi (b)}\ }\textbf {\bibinfo {volume} {252}},\ \bibinfo
  {pages} {828--832} (\bibinfo {year} {2015})}\BibitemShut {NoStop}%
\bibitem [{\citenamefont {Koksal}\ \emph {et~al.}(2018)\citenamefont {Koksal},
  \citenamefont {Tanen}, \citenamefont {Jena}, \citenamefont {Xing},\ and\
  \citenamefont {Rana}}]{koksal_opop}%
  \BibitemOpen
  \bibfield  {author} {\bibinfo {author} {\bibfnamefont {O.}~\bibnamefont
  {Koksal}}, \bibinfo {author} {\bibfnamefont {N.}~\bibnamefont {Tanen}},
  \bibinfo {author} {\bibfnamefont {D.}~\bibnamefont {Jena}}, \bibinfo {author}
  {\bibfnamefont {H.~G.}\ \bibnamefont {Xing}}, \ and\ \bibinfo {author}
  {\bibfnamefont {F.}~\bibnamefont {Rana}},\ }\href {\doibase
  10.1063/1.5058164} {\bibfield  {journal} {\bibinfo  {journal} {Applied
  Physics Letters}\ }\textbf {\bibinfo {volume} {113}},\ \bibinfo {pages}
  {252102} (\bibinfo {year} {2018})}\BibitemShut {NoStop}%
\end{thebibliography}%

\end{document}